\documentclass[12pt,preprint]{aastex}
\usepackage{amsmath} 
\shorttitle{Arcs as a broken ring.}
\shortauthors{C. Alard}
\begin{document}
\title{Gravitational arcs as a perturbation of the perfect ring.}
\author{C. Alard}
\affil{Institut d'Astrophysique de Paris, 98bis boulevard Arago, 75014
  Paris.}
\email{alard@iap.fr}
\begin{abstract}
The image of a point situated at the center of a circularly symmetric potential
is a perfect circle. The perturbative effect of non-symmetrical potential 
terms is to displace and break the perfect circle. These 2 effects, displacement
and breaking are directly related to the Taylor expansion of the perturbation
at first order on the circle. The numerical accuracy of this perturbative approach
is tested in the case of an elliptical potential with a core radius. The contour
of the images and the caustics lines are well re-produced by the perturbative approach.
These results suggests
that the modeling of arcs, and in particular of tangential arcs
 may be simplified by using a general perturbative
representation on the circle. An interesting feature of the perturbative approach,
is that the equation of the caustic line depends only on the values on the circle 
of the lens displacement field along the $\theta$ direction. 
\end{abstract}
\section{Introduction.}
Since the discovery of gravitational arcs in clusters of galaxies 
by Linds \& Petrosian (1986) and subsequently by Soucail {\it et al.} (1987),
the observation and study of arcs has developed considerably, and is now
becoming an essential tool in astrophysics. Arcs provides a wealth of 
information about the mass distribution in clusters of galaxies 
(see for instance: Comerford {\it et al. 2006}, \& Broadhurst {\it el al. 2006}).
However, the mass distribution of the astrophysical lenses is
complex and involves a large parameter space
which is difficult to explore. Thus, the derivation of a simplified perturbative theory
that is able to re-produce the general features of gravitational arcs
is an interesting tool to help understanding complex gravitational lenses.
\section{Basic ideas}
Let's assume that the projected density of the lens $\rho_s$ is
 circularly symmetric and centered at the origin. Let's also assume
that the lens is dense enough to reach critical density at a given 
radius $R_E$. Under such hypothesis, the image by the lens of a point 
source placed at the origin will be a perfect ring. We are now 
interested in small perturbations of this perfect ring. There are 
two type of perturbations to the point source perfectly
aligned in a circularly symmetric potential: first the source may
 not be perfectly at the center, and second
the potential may not be perfectly circular. Let's be more specific,
 in polar coordinates, the lens equation writes:
\begin{equation} \label{eq_1}
 {\bf r_S} = \left({\bf r}
 -\frac{\partial \phi}{\partial r} \right) {\bf u_r} - 
 \left( \frac{1}{r} \frac{\partial \phi}{\partial \theta} \right) {\bf u_\theta}
\end{equation}
In the unperturbed case, the equation reads:
\begin{equation} \label{unpert}
r-\frac{d \phi_0}{dr} = 0
\end{equation}
Where $\phi_0$ is a function of $r$ only. Let's now perturb this
equation by introducing a small displacement of the source from the
origin: ${\bf r_s}$, and a non-circular
perturbation to the potential, $\psi$.  Note that the perturbation
on $r_S$ and the perturbation on $\phi$ are assumed to be of the same
order. The perturbation may be described
by the following formula:
\begin{equation} \label{perts}
\begin{cases}{\bf r_S} &= \epsilon \ {\bf r_s}  \\
\phi &= \phi_0 + \epsilon \ \psi
\end{cases}
\end{equation}
Here $\epsilon$ is a small number: $\epsilon<<1$. Given a position $(r_s)$ for the source,
the image positions $(r,\theta)$, can be obtained by solving equation
(\ref{eq_1}). However solving directly may prove impossible in the 
general case. It is easier to find a perturbative solution by
inserting Eq. (\ref{perts}) in Eq. (\ref{eq_1}). 
Assuming that $\epsilon$ is small, the perturbation will introduce
a deviation from the perfect circle that will be  of
order $\epsilon$. An interesting feature of the un-perturbed solution
is that the image of a single point at the origin is a full circle,
which covers all range in $\theta$. Thus whatever the position
$\theta$ of the perturbed solution, there is always a point at the
same $\theta$ in the un-perturbed solution. However, the point in the
un-perturbed solution will be located at a slightly different radius
 $r$. The response to the perturbation on $r$ may be written, $r=R_E+
\epsilon \ dr$. For convenience, it is always possible to re-scale 
the coordinate system, so that the Einstein radius is exactly equal to unity,
in this case, the perturbation on $r$ is: 
\begin{equation} \label{r_def}
r = 1 + \epsilon \ dr
\end{equation}
To summarize, we must solve perturbatively Eq. (\ref{eq_1}), by
expanding around the un-perturbed solution $(1, \theta)$ for small
values of $\epsilon$. Note that this requires to expand the potential
at $r=1$. Using Eq. (\ref{perts}), the Taylor expansion of $\phi$
may be written:
\begin{equation} \label{phi_pert}
\phi = \phi_0 + \epsilon \psi = \sum_{n=0}^{\infty} \left [ C_n +
  \epsilon \ f_n(\theta) \right] \ (r-1)^n
\end{equation}
Where we define,
\begin{equation} \label{C_def}
  C_n = \frac{1}{n!} \left[ \frac{d^n \phi_0}{ d^n r} \right]_{(r=1)} 
\end{equation}
And:
\begin{equation} \label{fn_def}
 f_n(\theta) =  \frac{1}{n!} \left[ \frac{\partial^n \psi}{ \partial^n r}
 \right]_{(r=1)}
\end{equation}
 It is now possible to expand each side of Eq. (\ref{eq_1}) in series
 of $\epsilon$, in the vicinity of the un-perturbed solution. By 
 inserting Eq. (\ref{phi_pert}), and  Eq. (\ref{r_def}) in
 Eq. (\ref{eq_1}), the response $dr$ to the perturbation defined in
 Eq. (\ref{perts}) can be estimated to the first order in $\epsilon$.
$$
 \epsilon \ {\bf r_s} = (1+\epsilon \ dr - C_1 - \epsilon \ 2 \ C_2 \ dr -
 \epsilon \ f_1 ) \ {\bf u_r} -\epsilon \ \frac{\partial f_0}{\partial
    \theta} \ {\bf u_{\theta}}
$$
Note that Eq. (\ref{unpert}) implies that $C_1=1$, consequently:
\begin{equation} \label{final_pert_eq}
 {\bf r_s} = \left( \kappa \ dr - f_1 \right) \ {\bf u_r} - 
 \frac{\partial f_0}{\partial \theta} \ {\bf u_{\theta}}
\end{equation}
\indent With: $\kappa=1 -2 \ C_2$.
\section{Reconstruction of images.}
\subsection{Circular source contours.}
Let's consider a circular contour on a source with center ${\bf r_{0}}$
 radius, $R_0$. The equation for this contour is:
\begin{equation} \label{circle}
 {\bf \left(r_s-r_0\right)^2} = R_0^2
\end{equation}
Note that effect of the translation by the vector ${\bf r_0}=(x_0,y_0)$ can be taken into account
by re-defining $f_0$, and $f_1$ in Eq. (\ref{dr_eq}):
\begin{equation} \label{redef_eq}
 \begin{cases}
   {\bf r_s} = \left( \kappa \ dr - \bar f_1 \right) \ {\bf u_r} - 
 \frac{\partial \bar f_0}{\partial \theta} \ {\bf u_{\theta}} \\
 \bar f_i = f_i+x_0 \ \cos \theta+ y_0 \ \sin \theta \ \ \ \ i=0,1 \\
 \end{cases}
\end{equation}
The image by the lens of this contour can be obtained using
Eq's. (\ref{circle}) and (\ref{redef_eq}):
\begin{equation} \label{lensed_circle}
 R_0^2 = \left(\kappa \ dr \ -\bar f_1 \right)^2 + \left( \frac{\partial \bar f_0}{\partial \theta}  \right)^2
\end{equation}
 Solving Eq. (\ref{lensed_circle}), the
2 following solutions for $dr$ can be obtained:
\begin{equation}  \label{dr_eq}
 dr = \frac{1}{\kappa} \left[ \bar f_1  \pm \sqrt{R_0^2-\left( \frac{\partial \bar f_0}{\partial \theta} \right)^2} \right]
\end{equation}
In the above equation the condition for  image formation is:
$\Delta =  R_0^2-\left( \frac{\partial \bar f_0}{\partial \theta} \right)^2>0$. The mean position of the 2 contour lines
is  $\frac{\bar f_1}{\kappa}$, and the image width along the radial
direction is $\frac{2 \ \Delta}{\kappa}$.
\subsection{Elliptical source contours.}
Using Eq. (\ref{redef_eq}) the source Cartesian coordinates writes:
\begin{equation} \label{xs_ys}
\begin{cases}
  x_s=\left(\kappa \ dr-\bar f_1  \right) \cos \theta  + \frac{ d \bar f_0}{d \theta}  \sin \theta \\
  y_s= \left(\kappa \ dr-\bar f_1  \right) \sin \theta - \frac{ d \bar f_0}{d \theta}  \cos \theta \\
\end{cases}
\end{equation} 
The equation for an elliptical contour aligned with its main axis aligned with the coordinate
system is:
\begin{equation} \label{ellipse_cont}
 (1-\eta) \ x_s^2+ (1+\eta) \ y_s^2 = R_0^2
\end{equation}  
Using Eq.'s (\ref{ellipse_cont}) and (\ref{xs_ys}) one obtains:
\begin{equation} \label{ellipse_eq}
\begin{cases}
 dr=\frac{1}{\kappa} \left[f_1+\frac{\eta \ \sin 2 \theta}{S} \ \frac{d f_0}{d \theta} \pm \frac{\sqrt{R_0^2 \ S-(1-\eta^2) \ \left(\frac{d f_0}{d \theta}\right)^2}}{S} \right] \\
 S = 1-\eta \ \cos 2 \theta
\end{cases}
\end{equation}
\subsection{Numerical testing}
 The accuracy of Eq. (\ref{dr_eq}) has been tested for a source
 with a circular contour by direct comparison with numerical integration of the
lens equation using ray-tracing. For this application,
the potential proposed by Blandford \& Kochanek (1987) (hereafter,
 BK1987) was chosen. The
asymptotic isothermal form of the potential was selected. The potential reads:
\begin{equation}\label{pot_eq}
 \phi = s \ \sqrt{s^2+1} \left[ \sqrt{1+\frac{r^2 \ (1-\eta \cos 2\theta)}{s^2}}-1  \right]
\end{equation}
The ellipticity of the potential is controlled by the parameter $\eta$, BK1987 conclude that for physical
reasons  $\eta$ must be less than $0.2$. The circular, unperturbed potential $\phi_0$ corresponds
to $\eta=0$, in this case, at $r=1$, $\frac{\partial \phi}{\partial r}=1$ which implies that  $R_E=1$.
According to Eq's. (\ref{C_def}) and (\ref{fn_def}) the parameters of the perturbative expansion are:
\begin{equation} \label{f0_f1_c2}
\begin{cases}
\frac{d f_0}{d \theta}={\frac {\sqrt {{s}^{2}+1} \ \eta\sin  2 \theta }{\sqrt {
{s}^{2}+1-\eta \cos  2 \theta }}} \\
f_1={\frac {\sqrt {{s}^{2}+1} \left( 1-\eta\,\cos  2 \theta
  \right) }{\sqrt {{s}^{2}+1-\eta \cos  2 \theta }}}-1\\
\kappa = \frac{1}{1+s^2}\\
\end{cases}
\end{equation} 
Using equations (\ref{dr_eq}) combined with Eq. (\ref{f0_f1_c2}) it is possible to trace the 2 lines
that describe the image contour. The 2 solutions for $dr$ join at the contour edges. The result
is visible in Fig. (1). Note that the large tangential arc is well reproduced by the approximation
while there is some offset in the position of the inner image. The problem of the inner
image can be corrected using an iterative approach: the perturbative method gives a first
guess of the image position, at this initial position one may carry a local Taylor expansion
of the potential in order to find a better solution, and the procedure
may be iterated again. 
\subsection{Inverse modelling}
Given a system of arcs or arclets it is possible
to re-construct the perturbation field. Let's define
the arc system to reconstruct as a set of contours, with one contour
per image. For each contour a radial line of direction $\theta$ 
intersect the contour in 2 points, $r_1$ and $r_2$ (provided that
$\theta$ is in a suitable range).
It is simple to relate these 2 functionals to the fields, $f_1$ and
$f_0$ of the perturbative approach, in the case of a source with
circular contour (Eq. (\ref{dr_eq})):
\begin{equation} \label{inv_eq}
\begin{cases}
 \bar f_1 = \frac{\kappa}{2} \left( r_1+r_2  \right) +  C\\
 \frac{d \bar f_0}{d \theta} = \sqrt{R_0^2-\frac{\kappa^2}{4}(r_2-r_1)^2} \\ 
\end{cases}
\end{equation}
Where $C$ is a constant term.
Note that $\kappa$ is unknown, actually there is a fundamental
degeneracy in $\kappa$, only $f_1/\kappa$, $f_0/\kappa$, and $R_0/\kappa$
may be determined. By using inner of radial image in the iterative
approach mentioned in Sec. (3.3), in some case
it might be possible to break the degeneracy on $\kappa$. 
It is clear also that instead of circular contours,
one could have considered elliptical one, and Eq. (\ref{ellipse_eq})
may have been used. In such case, it is just a matter of where to put
the complexity, either in the source or the lens, and obviously the
judging criteria should be that as a whole the complexity is minimal.
\section{Caustics in the perturbative approach.}
 Caustics are singularities, which are defined by the simple property that the determinant of the Jacobian
Matrix $J$ is zero on the caustic lines.
$$
 J = \frac{\partial x_s}{\partial r} \frac{\partial y_s}{\partial \theta} -  \frac{\partial x_s}{\partial \theta} \frac{\partial y_s}{\partial r}= 0
$$
The calculation of the Jacobian is straightforward from Eq. (\ref{xs_ys}), it follows that:
\begin{equation} \label{J_dr}
  dr  = \frac{1}{\kappa} \left[ f_1+\frac{\partial^2 f_0}{\partial^2 \theta} \right]
\end{equation} 
Note that in this case a shift by a vector $\bf {r_0}$ does not have
to intorduced 
thus, $f_0=\bar f_0$, and $f_1=\bar f_1$.
Eq. (\ref{J_dr}) defines the critical lines. The caustics in the source plane can be obtained by inserting
Eq. (\ref{J_dr}) in Eq. (\ref{xs_ys}):
\begin{equation} \label{caustics}
\begin{cases}
 x_s = \frac{d^2 f_0}{d^2 \theta} \cos \theta + \frac{d f_0}{d \theta} \sin \theta \\
 y_s =  \frac{d^2 f_0}{d^2 \theta} \sin\ \theta - \frac{d f_0}{d \theta} \cos \theta \\
\end{cases}
\end{equation} 
Not that Eq. (\ref{caustics}) line depends only on $f_0$,
which is directly related to the multipole expansion of the potential
on the circle. The multipole expansion has also the advantage to
relate directly $f_0$ to the density by the means of the coefficients
of the multipole expansion at $r=1$. Turing now to a numerical application,
by inserting the expression of $f_0$ given in Eq. (\ref{f0_f1_c2}) in
Eq. (\ref{caustics}) a parametric equation
of the caustic line is obtained, the result is presented in Fig. (2). The
perturbative calculation of the caustic curve is accurate, this suggest that this
approximation may be used to derive general results on caustics. A simple
result is that:
\begin{equation} \label{re_eq}
 r_s^2 = \left(\frac{d f_0}{d \theta}\right)^2+ \frac{d^2 f_0}{d^2 \theta}
\end{equation}
Eq. (\ref{re_eq}) may be integrated other $\theta$, the second term $\frac{d^2 f_0}{d^2 \theta}$ due
to periodicity, it follows that the typical size of the caustics is
directly related to the variance of $\frac{d f_0}{d \theta}$. This result demonstrates that the
caustics cross-section, are closely related to the deviations of the
potential to circular symmetry, which is confirmed by the numerical analysis of
Meneghetti {\it et al.} (2007).
\begin{figure} \label{fig:plot_1}
\plotone{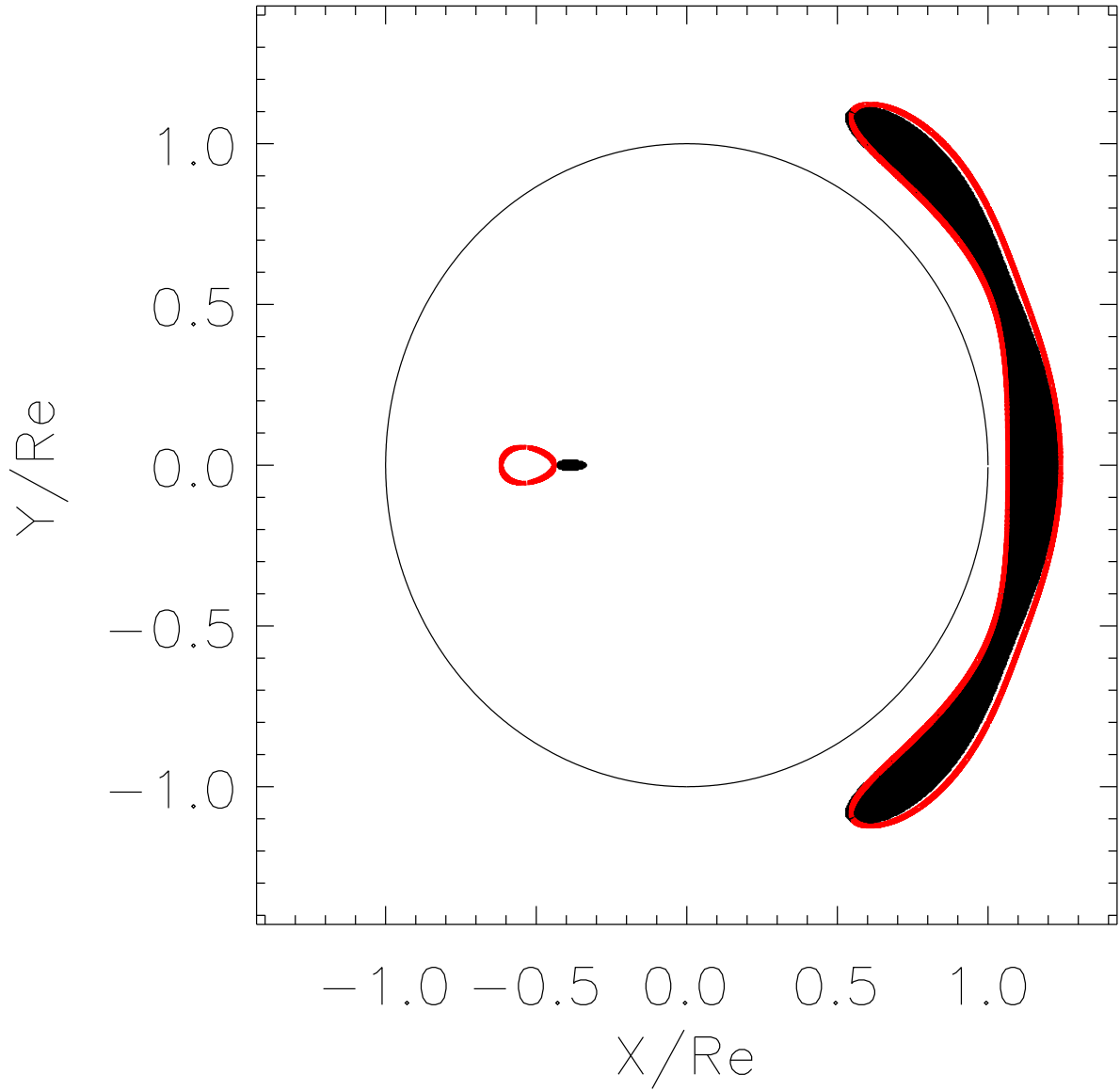}
\caption{A circular source of radius $R_0=\sqrt{2}/20 \ R_E$ is placed at  $[x_0=\frac{R_E}{5},y_0=0]$
 from the potential center along the X axis. The core radius of the potential is $s=\frac{R_E}{2}$, and the
ellipticity is $\eta=\frac{2}{10}$. The dark areas corresponds to the numerical solution
obtained by ray tracing. The outer contour of the source is reconstructed using Eq's (\ref{dr_eq}) and
(\ref{f0_f1_c2}) and super-imposed (as a red line) on the ray tracing solution. The red contour is close to
the outer contour of the ray-tracing solution for the tangential image. Due to the 
strong non-linearity of the potential near the center, the approximation is not as good for the inner image,
although as mentioned in Sec. (3.3) this may be corrected using an iterative approach.}
\end{figure}
\begin{figure} \label{plot_2}
\plotone{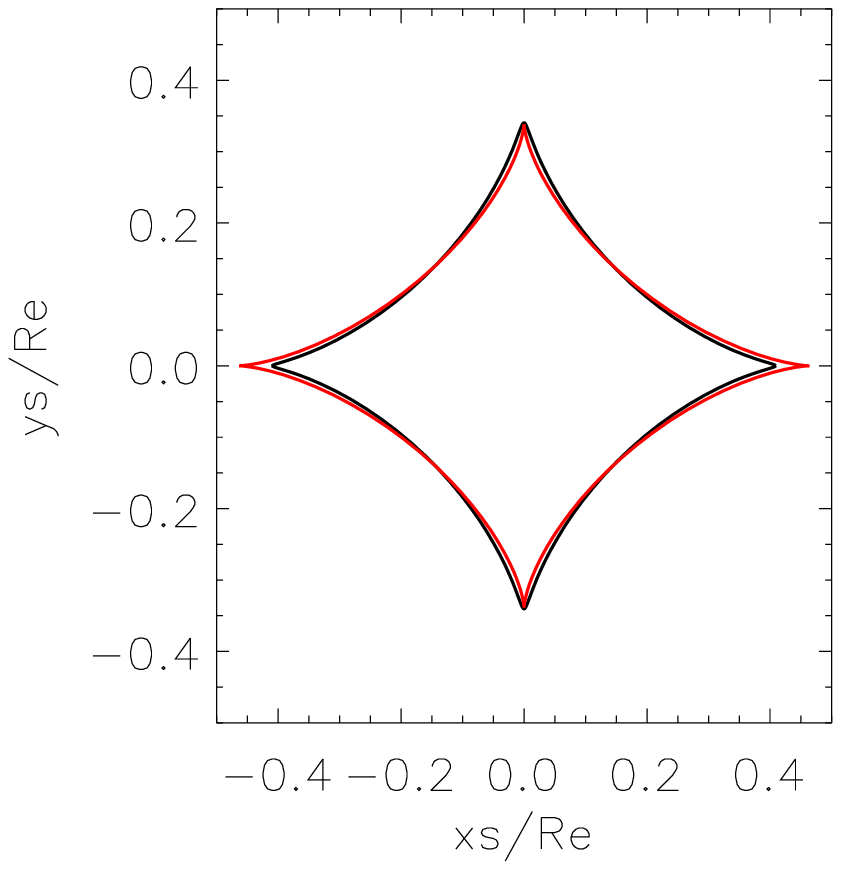}
\caption{Caustics of the BK1987 potential, the perturbative solution is plotted in black
 next to the numerical solution of the system of equations without approximations. Note
the closeness of the 2 solutions.}
\end{figure}
\acknowledgements{The author would like to thank J.P. Beaulieu and
  S. Colombi for reading this paper.}

\end{document}